\documentstyle[12pt]{article}

\newcommand{\be}{\begin{equation}}
\newcommand{\ee}{\end{equation}}

\def\psnormal{\textwidth=16cm\textheight=21.5cm
          \oddsidemargin=0.5cm\evensidemargin=0cm
          \topmargin=0cm\parindent=1cm}
\psnormal

\begin{document}
\pagestyle{empty}

\hspace{3cm}

\vspace{-0.6cm}
\rightline{{ CERN--TH.6835/93}}
\rightline{{IEM--FT--70/93}}

\vspace{0.6cm}
\begin{center}
{\bf {\large O}NE--{\large L}OOP
{\large A}NALYSIS OF THE {\large E}LECTROWEAK {\large B}REAKING
IN {\large S}UPERSYMMETRIC {\large M}ODELS
AND THE {\large F}INE--{\large T}UNING
{\large P}ROBLEM}
\vspace{0.8cm}

B. de CARLOS${}^*$ and J.A. CASAS${}^{**,*}\;$
\vspace{0.8cm}

${}^{*}$ Instituto de Estructura de la Materia (CSIC),\\
Serrano 123, 28006--Madrid, Spain
\vspace{0.4cm}

${}^{**}$ CERN, CH--1211 Geneva 23, Switzerland
\vspace{0.4cm}

\end{center}

\centerline{\bf Abstract}
\vspace{0.3cm}

\noindent
We examine the electroweak breaking mechanism in the minimal
supersymmetric standard model (MSSM) using the {\em complete}
one-loop effective potential $V_1$. First, we study
what is the region of the whole MSSM parameter
space (i.e. $M_{1/2},m_o,\mu,...$) that leads to a succesful
$SU(2)\times U(1)$ breaking with an acceptable top quark mass.
In doing this it is observed that all the one-loop corrections
to $V_1$ (even the apparently small ones) must be taken into account
in order to get reliable results. We find that
the allowed region of parameters is considerably enhanced
with respect to former "improved" tree level results.
Next, we study the fine-tuning problem associated
with the high sensitivity of $M_Z$ to $h_t$ (the top Yukawa coupling).
Again, we find that this fine-tuning is appreciably smaller once
the one-loop effects are considered than in previous tree level
calculations.
Finally, we explore the ambiguities and limitations of the ordinary
criterion to estimate the degree of fine-tuning.
As a result of all this, the upper bounds on the MSSM parameters,
and hence on the supersymmetric masses, are substantially raised,
thus increasing the consistency between supersymmetry
and observation.

\vspace{0.3cm}
\begin{flushleft}
{CERN--TH.6835/93} \\
{IEM--FT--70/93} \\
{March 1993}
\end{flushleft}
\psnormal
%
%

\newpage
\pagestyle{plain}
\pagenumbering{arabic}
\section{Introduction}

Precision LEP measurements give a strong support [1] to the expectations of
supersymmetric (SUSY) [2] grand unification [3]. Namely, the two loop
calculation indicates that the gauge coupling constants of the standard model
seem to be unified\footnote{This unification does not necessarily
require a GUT. In particular, in superstring theories all the gauge
couplings are essentially the same at tree level [4] even in the
absence of a grand unification group. This also avoids unwanted
consequences of GUT theories.} at $M_X\sim 10^{16}\ GeV$ with a value
$\alpha_X\sim 1/26$, provided the average mass of the new supersymmetric
states lies in the range [100 GeV, 10 TeV].

This calculation has been refined in a recent paper by Ross and Roberts
[5] in which the various supersymmetric thresholds were appropriately
taken into account. This was done in the context of the minimal
supersymmetric standard model (MSSM), which is characterised by the
Lagrangian
\begin{eqnarray}
{\cal L} = {\cal L}_{{\mathrm {SUSY}}}+{\cal L}_{{\mathrm {soft}}}\;\;.
\label{L}
\end{eqnarray}
Here ${\cal L}_{{\mathrm {SUSY}}}$ is the supersymmetric Lagrangian derived
from the observable superpotential $W_{{\mathrm {obs}}}$, which includes
the usual Yukawa terms $W_Y$ and a mass coupling $\mu H_1H_2$
between the two Higgs doublets $H_1$, $H_2$. ${\cal L}_{{\mathrm{soft}}}$
at the unification scale $M_X$ is given by
\begin{eqnarray}
{\cal L}_{{\mathrm{soft}}} = -m_{o}^2 \sum_\alpha |\phi_\alpha|^2
-\frac{1}{2}M_{1/2} \sum_{a=1}^3 \bar\lambda_a\lambda_a
-\left(Am_{o}W_Y + Bm_{o}\mu H_1H_2
 \ +\ \mathrm{h.c.}\right)
\label{Lsoft}
\end{eqnarray}
where $m_{o}$ and $M_{1/2}$ are the (common) supersymmetry soft breaking
masses (at $M_X$) for all the scalars $\phi_\alpha$ and gauginos $\lambda_a$
of the theory, and $A$ and $B$ parametrize the (common) couplings of
the trilinear and bilinear scalar terms. In this framework the physical
spectrum of supersymmetric masses depend on the particular choice
of the MSSM parameters
\begin{eqnarray}
m_o, M_{1/2},\mu,A,B,h_t
\label{softpar}
\end{eqnarray}
where $h_t$ is the top Yukawa coupling\footnote{These are the
parameters, together with the gauge
couplings , that enter in the renormalization group equations for
the masses. The influence of the bottom and tau Yukawa couplings is
negligible in most of the cases.}. Therefore, the requirement of
gauge unification constraints their ranges of variation.

These parameters are also responsible of the form of the Higgs scalar
potential and thus of the electroweak breaking process [6]. Requiring
the electroweak scale (i.e. $M_Z$) to be the correct one, together
with the presents bounds on $m_{top}$, Ross and Roberts further
restricted the allowed space of these parameters. Finally, these
authors imposed the absence of fine--tuning in the value of $h_t$ (the
parameter to which $M_Z$ is more sensitive\footnote{The sensitivity
of $M_Z$ to other independent parameters has been analyzed in ref.[7].})
for a successful
electroweak breaking, by demanding [7] $c\stackrel{<}{{}_\sim}10$
in the equation
\begin{eqnarray}
\frac{\delta M_Z^2}{M_Z^2} = c \frac{\delta h_t^2}{h_t^2}
\label{c}
\end{eqnarray}
where the value of $c$ depends on the values
of all the independent parameters listed in eq.(\ref{softpar}) (which
also determine the supersymmetric masses). As a consequence, they found
$m_o, \mu, M_{1/2}\stackrel{<}{{}_\sim}200\ $GeV (leading to typical
supersymmetric masses $\stackrel{<}{{}_\sim}500\ $GeV).
In fact, this turns out
to be the strongest constraint on the supersymmetric mass scale,
stronger than the requirement of gauge unification.

The analysis of ref.[5] of the electroweak breaking process and the
corresponding $h_t$-fine-tuning problem was performed by using the
renormalization improved tree level potential $V_o(Q)$, i.e. the
tree level potential in terms of the renormalized parameters at the
scale $Q$. However, as was shown in ref.[8], one expects the
effect of the one-loop contributions to be important. Consequently,
the analysis should be re-done using the whole one-loop effective
potential. This is the main goal of this paper.

In section 2 we study what is the region of the whole MSSM parameter
space (eq.(\ref{softpar})) leading to a correct  $SU(2)\times U(1)$ breaking
(this means a correct value for $M_Z$ and $m_{top}$
without color and electric charge breakdown).
The comparison with the results of the "renormalization improved"
tree level potential $V_o$ [5] shows that the one-loop corrections
enhance (and also displace) this allowed region.
As a by--product, we show that the (very common) approximation
of considering only the top and stop contribution (disregarding
the $\tilde t_L-\tilde t_R$ mixing) to the one-loop effective potential
is not reliable for analyzing the electroweak breaking mechanism.
In section 3 we analyze the above mentioned fine-tuning problem,
showing that, once the one-loop contributions are taken into account,
it becomes considerably softened. In addition to this, we study the
limitations and ambiguities of the ordinary criterion (\ref{c})
to estimate the fine-tuning problem. Although in the MSSM it turns out
to be a sensible criterion (which is not a general fact), it should
be considered as a rather qualitative one, thus the upper bound on
$c$ should be conservatively relaxed, at least up to
$c\stackrel{<}{{}_\sim}20$. As a consequence of all this, the
upper bounds on the MSSM parameters and on the supersymmetric masses
are pushed up from the "renormalized improved" tree level results.
This is relevant, of course, for the expectations of experimental
detection of SUSY. Finally, we present our conclusions in section 4.

\section{Radiative electroweak breaking}

In the MSSM the part of the tree-level potential along the neutral
components of the Higgs fields at a scale $Q$ is given by
\begin{eqnarray}
V_o(Q)=\frac{1}{8}(g^2+g'^2)\left(|H_1|^2-|H_2|^2\right)^2
+ m_1^2|H_1|^2 + m_2^2|H_2|^2 -m_3^2(H_1H_2+\mathrm{h.c.})
\;\;,
\label{Vo}
\end{eqnarray}
where
\begin{eqnarray}
m_i^2=m_{H_i}^2+\mu^2\;,\;\;\;m_3^2=m_o\mu B
\label{emes}
\end{eqnarray}
with
\begin{eqnarray}
m_{H_i}^2(M_X)=m_o^2
\label{emes2}
\end{eqnarray}
In the usual calculations with just
the tree level potential $V_o(Q)$ (as in ref.[5]), this was minimized
at the $M_Z$ (or $M_W$) scale.

The one-loop effective potential is given by [9]
\begin{eqnarray}
V_1(Q)=V_o + \Delta V_1
\label{V1}
\end{eqnarray}
where
\begin{eqnarray}
\Delta V_1(Q)=\frac{1}{64\pi^2} Str\left[
{\cal M}^4\left(\log\frac{{\cal M}^2}{Q^2}-\frac{3}{2}\right)
\right]
\label{deltaV1}
\end{eqnarray}
depends on $H_1$, $H_2$ through the tree-level squared-mass matrix
${\cal M}^2$. In the expressions
(\ref{Vo},\ref{emes},\ref{V1},\ref{deltaV1}) all the parameters
are understood to be running parameters evaluated at the scale $Q$.
They can be computed by solving the standard renormalization group
equations (RGE's), whose form is well known [2], and taking into account
all the supersymmetric thresholds. The supertrace of
eq.(\ref{deltaV1}) runs over {\em all} the states of the theory.
This, in particular, amounts to determine the eigenvalues of the mass
mixing matrices of stops, charginos and neutralinos. Incidentally,
a simplification broadly used in the literature is to consider just
the top ($t$) and stop
($\tilde t$) contributions to (\ref{deltaV1}), disregarding also the
$\tilde t_L-\tilde t_R$ mixing. This can be a good approximation
for certain purposes (see e.g. ref.[10]), but, as will be shown
shortly, it is {\em not} when one is interested
in studying the $SU(2)\times U(1)$ breaking.
To be in the safe side the whole spectrum contribution must
be considered in eq.(\ref{deltaV1}).

In order to exhibit the implications of considering the whole
one-loop potential $V_1$ versus $V_o$, we have shown two examples
({\em a}) and ({\em b}) in fig.1. They are specified by the
the following initial values of the independent
parameters
\begin{eqnarray}
&(a)\hspace{1cm}&\hspace{1cm} m_o=\mu= 120 \ GeV, M_{1/2}=230 \ GeV,
A=B=0, h_t=0.207 \nonumber \\
&(b)\hspace{1cm}&\hspace{1cm} m_o=\mu=100 \ GeV, M_{1/2}=180 \ GeV,
A=B=0, h_t=0.250
\label{casosab}
\end{eqnarray}
The case ({\em a}) corresponds to one of the two models explicitly
expounded in ref.[5] (where it was called "X").
Although in
the $V_o$ approximation this model works correctly, once the one-loop
contributions are considered, we see that it does not even lead
to electroweak breaking (the same happens with the model that was
called "Z"). In the example ({\em b}) both $V_o$ and
$V_1$ yield electroweak breaking, but for completely different
values of $v_1\equiv \langle H_1\rangle$ and
$v_2\equiv \langle H_2\rangle$. In this case, $V_1$ predicts
electroweak breaking at the right scale, while $V_o$ does not.
The above-mentioned approximation of considering just the top
and stop contribution to $\Delta V_1$, which is also represented
in the figure, works better than $V_o$, but not enough to produce
acceptable results. Moreover, it is clear from the figure
that only the {\em whole} one--loop contribution
really helps to stabilize the values of $v_1, v_2$ versus variations
of $Q$ (they are essentially constant up to $O(\hbar^2)$ corrections).
In fact, they should evolve only via the (very small) wave
function renormalization effects, given by
\begin{eqnarray}
\frac{\partial \log v_1}{\partial \log Q}&=&\frac{1}{64\pi^2}
(3g_2^2+g'^2)
\nonumber \\
\frac{\partial \log v_2}{\partial \log Q}&=&\frac{1}{64\pi^2}
(3g_2^2+g'^2-12h_t^2)\;\;.
\label{v1v2Q}
\end{eqnarray}

There is a scale, that in ref.[8] was called $\hat Q$, at which
the results from $V_o$ and $V_1$ approximately coincide. At
this scale the one-loop contributions are quite small, in
particular the logarithmic factors, so $\hat Q$ represents a
certain average of all the masses. In the region around $\hat Q$
one expects, due to the smallness of the logarithms, that the
evaluation of one-loop effects is more reliable (see also
ref.[11]).

In the example depicted in fig.1b this consideration is not
very relevant, for $v_1$ and $v_2$ are essentially constant.
However, there are cases where $v_1(Q)$ and $v_2(Q)$ do not show
such a remarkable stability. This happens when the averaged
supersymmetric mass is much larger than $M_Z$, since this
leads to the appearance of large logarithms at $Q=M_Z$
(this fact has been stressed in ref.[11]). However, in the
region around $\hat Q$ (i.e. precisely where the calculation
is more reliable) $v_1(Q)$ and $v_2(Q)$ are {\em always}
stable. Thus we have used the following criterion: we
evaluate $v_1$ and $v_2$ at the $\hat Q$ scale and then
we calculate $v_1(Q)$ and $v_2(Q)$ via eq.(\ref{v1v2Q})
at any other scale. This is relevant at the time of calculating
physical masses. In particular $M_Z$ is given by
\begin{eqnarray}
\left. (M_Z^{\mathrm{phys}})^2\simeq \frac{1}{2}\left(g_2^2(Q)
+g'^2(Q)\right)\left[
v_1^2(Q)+v_2^2(Q)
\right]\;\right|_{Q=M_Z^{\mathrm{phys}}}
\label{Mz}
\end{eqnarray}
and similar expressions can be written for all the particles of
the theory.

Now we are ready to determine how the requirement of correct electroweak
breaking (i.e. $M_Z^{\mathrm{phys}}= M_Z^{\mathrm{exp}}$) puts
restrictions in the space of parameters. "Correct electroweak
breaking" of course means $M_Z^{\mathrm{phys}}= M_Z^{\mathrm{exp}}$,
where $M_Z$ is given by (\ref{Mz}).
In addition, other physical requirements must be satisfied. Namely,
the scalar potential must be bounded from below [2], color and
electric charge must remain unbroken [2], and the top mass must
lie within the LEP limits ($100\;GeV\stackrel{<}{{}_\sim}
m_{top}\stackrel{<}{{}_\sim}160\;GeV$).
Following a similar presentation
to that of ref.[5], the results of the analysis for $A=B=0$ (at $M_X$)
and for various initial values of $|\mu_o/m_o|$ are shown in fig.2.
The value of $\alpha_3(M_Z)$
necessary to achieve unification of the couplings was calculated
in ref.[5] at the two-loop order and is also represented in the figure.
We have also evaluated the effect of varying the $A$ and $B$ parameters,
as is illustrated in fig.3.
The effect of the one loop contribution is to enhance
and displace the
region of allowed parameters appreciably. In order to facilitate
the comparison we have reproduced in fig.4 the $V_o$ results [5]
and the one-loop results for for the case of fig.2c
(i.e. $m_o/\mu_o=1, A=B=0$), which is a representative one.

\section{The fine-tuning problem}

As was pointed out in ref.[5], $h_t$ is the parameter to which the
value of $M_Z$ is more sensitive. This sensitivity is conveniently
quantified by the $c$ parameter defined in eq.(\ref{c}). We have
represented the values
$c$ for the representative case of fig.4.
A good
parameterization of the value of $c$ is
\begin{eqnarray}
c\simeq \frac{1}{M_Z^2}\left[1.08\ M_{1/2}^2+0.19\ (m_o^2+\mu_o^2)\right]
\label{parc}
\end{eqnarray}
The high influence of $M_{1/2}$ on the value of $c$ compared to
that of $m_o$ and $\mu_o$ comes from the fact that scalar masses can be
very high, even if they are vanishing at tree level, due
to the gaugino contribution in the RGE's, but not the other
way round. The
tree level results [5] are also given to facilitate the
comparison\footnote{We reproduce here the values of $c$ for $V_o$ as
given in ref.[5], though our calculation gives slightly different values.}.
The sensitivity of $M_Z$ to $h_t$ turns out to be substantially smaller
with the complete one-loop effective potential than with the $V_o$
approximation. If, following ref.[5], we demand now
$c\stackrel{<}{{}_\sim}10$ as the criterion to
avoid the fine-tuning in $h_t$, this selects a region of acceptable
SUSY parameters that can easily be read from fig.4. Notice that this
region is noticeably larger than the corresponding one obtained
from $V_o$. This is a consequence of the lower sensitivity of $M_Z$
to $h_t$ {\em and} the larger region of parameters giving a correct
value of $M_Z$ (see section 2) when one uses the entire one-loop
effective potential $V_1$. Accordingly, the one-loop
contributions tend to make less "critical" the electroweak breaking
process in supersymmetric models.

We would also like to make some comments on the criterion usually
followed to parameterize the fine-tuning problem, i.e.
$c\stackrel{<}{{}_\sim}10$ in eq.(\ref{c}). First of all,
to some extent this procedure is ambiguously defined, since it
depends on our definition of the independent parameters and
the physical magnitude to be fitted. For example, if we replace
$M_Z^2$ by $M_Z$ in eq.(\ref{c}), then the corresponding values
of $c$ (represented in fig.3) are divided by two. Second,
notice that if for a certain choice of the supersymmetric
parameters ($m_o, M_{1/2},\mu, A, B$), the value of $c$
turned out to be high for most of the possible values of
$h_t$ (or equivalently $M_Z$), then we would arrive to the
bizarre conclusion that {\em any} value of $h_t$ leads to a
fine-tuning!\footnote{This would happen, for instance, if the
hypothetic theoretical relation between $M_Z$ and $h_t$
were $M_Z\sim \exp\{Ch_t\}$ with $|Ch_t|>10$.}. This is so because
the "standard" criterion
of eq.(\ref{c}) measures the {\em sensitivity} of $M_Z$ to $h_t$
rather than the degree of fine-tuning. In order for eq.(\ref{c})
to be a sensible quantification of the fine-tuning it should be
required $c\sim 1$ for most of the $h_t$ values. To check this,
we have represented in fig.5 $M_Z$ versus $h_t$ for
a typical example ($m_o=\mu=M_{1/2}=500\;GeV,A=B=0$). We
see that, indeed, for most of the $h_t$ values the sensitivity
of $M_Z$ to $h_t$ is small. Hence, the parameterization of the
fine-tuning by the value of $c$ in eq.(\ref{c}) is meaningful.
A natural value for $M_Z$ under
these conditions would be $M_Z\sim 1\; TeV$.\footnote{Notice, however,
that if we restrict the range of variation of $h_t$ so that
$100\; GeV<m_{top}<160\; GeV$, then $c>10$ in the entire
"allowed" region of $h_t$.} Nevertheless, this shows that it
is dangerous to assume that $c$ is an exact measure of the
degree of fine-tuning. It is rather a sensible, but qualitative
one. In fact, a precise evaluation of the degree of fine-tuning
would require a knowledge of what are the actual independent
parameters of the theory and what is the supergravity breaking
mechanism (for an example of this see ref.[12]).

All the previous considerations suggest that the upper limit
$c\stackrel{<}{{}_\sim}10$ in the measure of the allowed
fine-tuning should be conservatively relaxed, at least
up to $c\stackrel{<}{{}_\sim}20$. We see from fig.4 that this
implies
\begin{eqnarray}
m_o, \mu\stackrel{<}{{}_\sim}650 \ GeV,\;\;\; M_{1/2}
\stackrel{<}{{}_\sim}400 \ GeV
\label{cotassoft}
\end{eqnarray}
In order to see what are the corresponding
upper limits on the supersymmetric masses,
we have explicitly given the mass spectrum (including also
the small contributions coming from the electroweak breaking)
in Table 1 for the two "extreme" cases
labelled as $X_1$ and $X_2$ in fig.4. Note that these
two cases are close to the $c=20$ line and to the upper and
lower limits on the top quark mass. From these extreme examples
we see that, roughly speaking, the bounds on the most relevant
supersymmetric particles are
%
\begin{equation}
\begin{array}{cc}
\mathrm{Gluino}\hspace{2cm}& M_{\tilde g}\stackrel{<}{{}_\sim}1100\; GeV \\
\mathrm{Lightest \ chargino}\hspace{2cm}& M_{\chi^{\pm}}\stackrel{<}{{}_\sim}
250\;GeV \\
\mathrm{Lightest \ neutralino}\hspace{2cm}& M_{\lambda}\stackrel{<}{{}_\sim}
200\;GeV \\
\mathrm{Squarks}\hspace{2cm}& m_{\tilde q}\stackrel{<}{{}_\sim}900\;GeV \\
\mathrm{Sleptons}\hspace{2cm}& m_{\tilde l}\stackrel{<}{{}_\sim}450\;GeV
\end{array}
\label{cotas}
\end{equation}
These numbers are substantially higher than those obtained in ref.[5]
from $V_o$,
and summarize the three main results obtained in this paper: {\em i)}
The region of parameters giving a correct
electroweak breaking is larger when one uses the entire one-loop
effective potential $V_1$ than with $V_o$ (see section 2),
{\em ii)} The corresponding sensitivity of $M_Z$
to the value of $h_t$ is smaller and {\em iii)} The highest acceptable
value of $c$ (see eq.(\ref{c})) must be conservatively relaxed for
the above explained reasons. The most important conclusion at
this stage is that the supersymmetric spectrum is not necessarily
close to the present experimental limits, though the future
accelerators (LHC, SSC) should bring it to light. It is also remarkable
that the $\tilde t_L-\tilde t_R$ splitting can be very sizeable
in many scenarios. Let us finally note that there are considerable
radiative corrections to the lightest Higgs mass coming from the
top-stop splitting [11], which have not been included in Table 1.

\section{Conclusions}

We have studied the electroweak breaking mechanism in the minimal
supersymmetric standard model (MSSM) using the {\em complete}
one-loop effective potential $V_1=V_o+\Delta V_1$ (see
eqs.(\ref{Vo},\ref{V1},\ref{deltaV1}). We have focussed the attention
on the allowed region of the parameter space leading to a correct
electroweak breaking, the fine-tuning problem and the upper bounds
on supersymmetric masses.

As a preliminary, we showed that some common approximations, such
as considering only the top and stop contributions to $\Delta V_1$
and/or disregarding the $\tilde t_L-\tilde t_R$ mixing, though
acceptable for other purposes, lead to wrong results for $SU(2)\times
U(1)$ breaking. In consequence, we have worked with the exact one-loop
effective potential $V_1$.

Next, we have examined what is the region of the whole MSSM parameter
space (i.e. the soft breaking terms $M_{1/2},m_o,A,B$ plus $\mu$
and $h_t$) that leads to a correct  $SU(2)\times U(1)$ breaking,
i.e. the correct value of $M_Z$, a value of $m_{top}$ consistent
with the observations and no color or electric charge breakdown.
A comparison with the results of the "renormalization improved"
tree level potential $V_o$ [5] shows that the one-loop corrections
enhance (and also displace) the allowed region of parameters. This,
of course, are good news for the MSSM.

Our following step has been to analyze the top-fine-tuning problem.
As it has been pointed out in ref.[5], $h_t$ (the top Yukawa coupling)
is the parameter to which $M_Z$ is more sensitive. Using the ordinary
criterion to avoid fine-tuning, i.e. $c\stackrel{<}{{}_\sim}10$
in the relation
\begin{eqnarray}
\frac{\delta M_Z^2}{M_Z^2} = c \frac{\delta h_t^2}{h_t^2}\;\;\;,
\label{c2}
\end{eqnarray}
strongly constraints the values of the MSSM parameters, leading to
upper bounds on $M_{1/2},m_o,\mu$, and thus on the masses of the new
supersymmetric states (gluino, squarks, charginos, etc.). This analysis
was performed in ref.[5] using the improved tree level potential $V_o$.
We find that the one-loop corrections substantially soften the degree
of fine-tuning. This, again, are good news for the MSSM.

Finally, we have explored what are the limitations of the ordinary
criterion (\ref{c2}) to parameterize the degree of fine-tuning.
We comment on its ambiguities and show a type of (hypothetical)
scenarios in which this criterion would be completely meaningless.
Fortunately, this is not the case for the MSSM and, thus, the $c$
parameter represents a sensible, but qualitative estimation
of the degree of fine-tuning. A precise and non-ambiguous
quantification of it can only be done once one knows the
supergravity breaking mechanism. In view of all this, we have
conservatively relaxed the acceptable upper bound for $c$
up to  $c\stackrel{<}{{}_\sim}20$.

As a summary of the results the one-loop contributions
{\em i)} enhance (and displace) the allowed
region of the MSSM parameters
{\em ii)} soften the fine-tuning associated with the top quark
(for large values of the MSSM parameters). These two facts
together with the fact that
{\em iii)} the upper bound on $c$ should be conservatively relaxed,
push up the upper bounds on the MSSM parameters obtained
from the former $V_o$ analysis {\em and} the corresponding
upper bounds on supersymmetric masses. This is reflected in Table 1
for two "extreme" cases and in eq.(\ref{cotas}). Our final
conclusion is that the supersymmetric spectrum is not necessarily
close to the present experimental limits, though the future
accelerators (LHC, SSC) should bring it to light.

\vspace{2cm}

\noindent{\bf ACKNOWLEDGEMENTS}

\vspace{0.3cm}
We thank C. Mu\~noz and J.R. Espinosa for very useful discussions
and suggestions. We also thank O. Diego, F. de Campos and P.
Garc\'{\i}a-Abia
for their invaluable help with the computer.
The work of B.C.
was supported by a Comunidad de Madrid grant.

\newpage


\renewcommand{\arraystretch}{1.2}

\begin{table}
\underline{\bf TABLE 1}

\begin{tabular}{|ccc|}
\hline
\multicolumn{3}{|c|}{Parameters (initial values)} \\ \hline
$M_{1/2}$ (GeV) & 300 & 400 \\
$m_o$ (GeV) & 400 & 200 \\
$\mu$ (GeV) & 400 & 200 \\
$h_t$ & 0.618 & 0.254 \\
$A,B$ & 0 & 0 \\ \hline
\multicolumn{3}{|c|}{Masses of Gluino, Charginos and Neutralinos
(in GeV)} \\ \hline
$\tilde g$  & 837 & 1124 \\
$\chi_1^{\pm}$  & 407 & 376 \\
$\chi_2^{\pm}$  & 243 & 226 \\
$\lambda_1$  & 172 & 169 \\
$\lambda_2$  & 242 & 371 \\
$\lambda_3$  & 408 & 236 \\
$\lambda_4$  & 387 & 255 \\
\hline
\multicolumn{3}{|c|}{Masses of Squarks (in GeV)} \\ \hline
$\tilde u_L,\tilde c_L;\ \tilde d_L,\tilde s_L $  & 785; 789 & 922; 925 \\
$\tilde u_R,\tilde c_R$  & 766 & 888 \\
$\tilde d_R,\tilde s_R,\tilde b_R$  & 762 & 885 \\
$\tilde t_L,\tilde b_L$  & 827, 698 & 1055, 881 \\
$\tilde t_R$  & 410 & 560 \\ \hline
\multicolumn{3}{|c|}{Masses of Sleptons and Higgses (in GeV)} \\ \hline
$\tilde l_L,\tilde l_R$  & 476, 431 & 372, 256 \\
$h^o,H^o$  & 91, 547 & 91, 353 \\
$H^{\pm}$  & 553 & 362 \\
$A^o$  & 547 & 353 \\ \hline
\end{tabular}
\caption{Masses of the supersymmetric states for the two solutions
(called $X_1$ and $X_2$ in fig.4) with $m_{top}=163, 109$ respectively.
All the masses are given at the $M_Z$ scale.}
\end{table}

\newpage
\pagestyle{plain}
\pagenumbering{arabic}
\setcounter{page}{12}

\vspace{0.3cm}
\noindent{\bf FIGURE CAPTIONS}

\begin{description}

\item[Fig.1] $v_1\equiv \langle H_1\rangle$, $v_2\equiv
\langle H_2\rangle$ versus the $Q$ scale between $M_Z$ and 2 TeV
(in GeV) for the cases
labelled as (a) and (b) in eq.(10). Solid lines: complete
one-loop results; dashed lines: "improved" tree level
results; dotted lines: one-loop results in the top--stop
approximation.

\item[Fig.2] Allowed values for the $M_{1/2}$, $m_o$ parameters
(in GeV) for different vales of $\mu_o$: $|\mu_o/m_o|=0.2,0.4,1,3$ in
(a), (b), (c), (d) respectively, and $A=B=0$. The solid lines
represent the value of $\alpha_3(M_Z)$ needed to achieved unification,
as calculated in ref.[5]. Dotted lines correspond to the
extreme values of $m_{top}$ (evaluated at the $M_Z$ scale):
$m_{top}=160,100$ GeV.

\item[Fig.3] The same as fig.2, but for different values of $A,B$:
$A=0,0,1,-1$, $B=0,1,0,0$ in
(a), (b), (c), (d) respectively, and $|\mu_o/m_o|=1$. In case (c), the
$m_{top}=160$ GeV line coincides with the $M_{1/2}=100$ GeV axis.

\item[Fig.4] The case $A=B=0$,
$|\mu_o/m_o|=0.2,0.4,1,3$  with the  "improved" tree level potential
$V_o$ (a) and the whole one-loop effective potential $V_1$ (b).
Diagonal lines correspond to the estreme values
of $m_{top}$, as were calculated by Ross et al. in ref.[5]:
$m_{top}=160,100$ GeV. Transverse
lines indicate constant values of $c$, defined in eq.(4).

\item[Fig.5] $M_Z$ versus $h_t$ for $M_{1/2}=m_o=\mu_o=500$ GeV,
$A=B=0$. The region of physical $M_Z$ amounts a fine-tuning in the value
of $h_t$.

\end{description}
\end{document}